\documentclass[11pt]{article}
\def \beq  {\begin{equation}}
\def \eeq  {\end{equation}}
\def \beqar {\begin{eqnarray}}
\def \eeqar {\end{eqnarray}}

\def\sqr#1#2{{\vcenter{\vbox{\hrule height.#2pt
\hbox{\vrule width.#2pt height#1pt \kern#1pt
\vrule width.#2pt}\hrule height.#2pt}}}}

\def\la {{\langle}}
\def\ra {{\rangle}}

\def\vf {{\varphi}}

\def\Tr {{\rm Tr}}

\def \dotA {{\dot A}}
\def \dotB {{\dot B}}

\def\bA {\bar{A}}

\def\bu {\bar{u}}

\def\bw {\bar{w}}

\def\del {\partial}

\def\D {{\cal D}}
\def\bz {{\bar{z}}}

\def\vf {{\varphi}}

\def\half{\textstyle{1\over 2}}

\begin{document}

\begin{titlepage}
\null\vspace{-62pt} \pagestyle{empty}
\begin{center}
\rightline{CCNY-HEP-04/9}
\rightline{August 2004}
\vspace{1truein} {\Large\bf
Multigluon amplitudes, ${\cal N} =4$ constraints and\\
\vskip .2in\noindent
the WZW model}\\

\vspace{.6in}
YASUHIRO ABE$~^{a}$, V. P.~NAIR$~^{b}$ and MU-IN PARK$~^{c}${\footnote{
E-mail addresses:
abe@sci.ccny.cuny.edu (Y. Abe), vpn@sci.ccny.cuny.edu (V.P. Nair),
muinpark@yahoo.com (M.I. Park)}}\\
\vskip .1in
$^{a,b}$ {\it Physics Department, City College of the CUNY\\
New York, NY 10031}\\
\vskip .1in
$^{c}$ {\it Department of Physics, POSTECH \\
Pohang 790-784, Korea}\\

\vspace{.8in}
\centerline{\large\bf Abstract}
\end{center}
Classical ${\cal N}=4$ Yang-Mills theory is defined by the superspace constraints.
We obtain a solution of a subset of these constraints and show that it leads to
the maximally helicity violating (MHV) amplitudes.
The action which leads to the solvable part of the
constraints is a Wess-Zumino-Witten (WZW) action
on a suitably $extended$ superspace.
The non-MHV tree amplitudes can also be expressed
in terms of this action.

\end{titlepage}

\pagestyle{plain} \setcounter{page}{2}

\section{Introduction}

The construction of multigluon scattering amplitudes in the
${\cal N}=4$ super Yang-Mills theory has attracted a lot of attention
recently.
The calculation of some of these amplitudes,
particularly the so-called maximally helicity violating (MHV) ones,
was carried out quite some time ago
\cite{pt}.
Although the intermediate steps of the calculation
were algebraically very complex,
the final results were surprisingly simple.
It was pointed out shortly afterward that the MHV amplitudes could be obtained
in terms of the current correlators of a Wess-Zumino-Witten (WZW)
theory and that there was a natural
interpretation of this in supertwistor space \cite{nair}.
(For some further developments along this direction, see
\cite{others}. For a discussion of twistor space, see
\cite{penrose}; for supertwistor space, see \cite{ferber}.)
Recently, Witten showed that there is a deep connection of these results
to string theory \cite{witten}.
First of all, the supertwistor
space ${\bf CP}^{3|4}$, as a supermanifold, is a
Calabi-Yau space, so that it is possible to have a string theory
with this target space. A topological
version of such a string theory, the so-called
topological $B$-model, can be constructed.
The MHV amplitude is the restriction
of a holomorphic function in ${\bf CP}^{3|4}$ to
a complex line.
This complex line can be interpreted as a $D$-instanton in the
string theory.
The correlators of the $B$-model on this line become WZW correlators,
reproducing the MHV amplitudes.
One of the key observations in \cite{witten}
was that the non-MHV amplitudes can be obtained
as the correlators of the $B$-model restricted to
algebraic curves of higher degree in ${\bf CP}^{3|4}$.
This seems to be true by direct verification of many
amplitudes \cite{list1}. Later, it was realized that one could perhaps
simplify even more \cite{witten3}. The amplitudes can be constructed
by considering the MHV amplitudes, with a suitable
off-shell continuation, as the basic vertices.
By connecting such vertices via propagators,
one can obtain all the gauge theory amplitudes.
This too seems to be born out by explicit calculations
carried out so far \cite{list2}. It is a remarkable result,
with all the tree amplitudes of the gauge theory obtained by a simple
set of rules in twistor language.

An alternative string theory which leads to the
same amplitudes
has been proposed by Berkovits \cite{berk}.
A number of other related works, including
ramifications of these results in string theory
are given in \cite{string}.

While these are remarkable developments, in this paper, we go back to the
well-known formulation of
supersymmetric gauge theories in terms of
gauge potentials in superspace. Generally, such gauge potentials
contain too many degrees of freedom, more than what is needed for
the physical fields. One can then impose a set of constraints obeyed by the
field strengths in superspace; these constraints can be solved in terms of
some unconstrained fields and the latter can be used for the construction
of the action for the theory.
However, in the case of the ${\cal N}=4$ Yang-Mills theory,
the constraints are too stringent and, in fact, imply the equations of motion
via the Bianchi identities \cite{sohnius}.
For the purpose of constructing an action
with manifest ${\cal N}=4$ supersymmetry, this is bad news since
we do not have
fields which are off the mass-shell. However, the good news is that
this property shows that the classical
equations of motion are equivalent to a set of first order
equations in the appropriate superspace. One could then hope, in a way similar to the
strategy for solving the first order self-duality (instanton) equations, that
the constraints of the ${\cal N}=4$ can be solved.
This is precisely what we attempt to do in this paper.
Our approach has similarities to the use of the holomorphic Chern-Simons theory
\cite{witten, popov}.
We introduce auxiliary bosonic variables to enlarge the space on which the constraints
are written. A `gauge transformation' in this enlarged space is then made to eliminate
some of the gauge potentials. The version of the constraints in the new gauge
are then solved with one additional simplifying assumption.
This leads to the formula for the MHV amplitudes.
The suggestion made in
\cite{witten3} can be incorporated in this language rather neatly.

In the next section, we set up the connection between the MHV amplitudes and the
WZW action. Section 3 is devoted to the solution of the constraints of
the ${\cal N}=4$ theory and the resulting $S$-matrix.
In section 4 we show how the non-MHV amplitudes, along the
lines of \cite{witten3}, can be phrased in our language.

\section{Multigluon amplitudes}

We start our discussion by writing the WZW action
in a form suitable for our purpose \cite{Witt:84}. As is well known, the WZW
action is related to the chiral Dirac determinant and so, taking
$A_{\bar{z}} (z,\bar{z})$'s
 in the fundamental representation of ${\rm SU(N)}$, we can write

\beqar
{\cal S}_{WZW}(M^\dagger ) &=& \Tr \log D_\bz - \Tr \log
\del_\bz = \Tr \log [1+ (\del_\bz )^{-1}
A_\bz ]\nonumber\\
&=& \sum_{n=2}^\infty {(-1)^{n+1}\over n}
\int {d^2z_1\over \pi} \cdots {d^2z_n \over \pi} \Tr\left[{A_\bz
(1) A_\bz (2) \cdots A_\bz (n) \over z_{12} z_{23} \cdots
z_{n1}}\right] ,
\nonumber\\
\label{cp1}
\eeqar
where $A_\bz = M^{\dagger -1} \del_\bz M^\dagger$, $D_\bz$ is the covariant
derivative $\del_\bz + A_\bz$, and
we have used the fact that the inverse
of $\del_\bz$ is given by $[\pi (z -z')]^{-1}$ and
$z_{mn} = z_m - z_n$. Also
$A_{\bar{z}} (n)$ denotes $A_{\bar{z}}
(z_n,\bar{z_n})$ and 
$d^2z$
is the real two-dimensional
volume element, equal to $ dz d\bz /(-2i)$, in the complex coordinates 
$z,\bar{z}$ for the Riemann surface.

The derivative of the action with respect to $A_\bz$ defines the
expectation value of the fermion current $J$
which minimally couples to $A_{\bar{z}}$; 
we can, therefore, express the above equation as a series of
current correlators,
\beq
\la J^{a_1}(1) J^{a_2}(2) \cdots
J^{a_n}(n)\ra = {(-1)^{n+1} \over n \pi^n}\left[ {\Tr(t^{a_1}
t^{a_2}\cdots t^{a_n}) \over z_{12} z_{23} \cdots z_{n1}}+ {\rm
permutations}\right].
\label{cp2} 
\eeq

We now introduce a spinor variable $u^A$, $A= 1,2$,
\beq
u = \left( \matrix{ \alpha\cr \beta\cr}\right).
\label{cp3}
\eeq
The complex projective space
${\bf CP}^1$ is defined by making
the identification $u \sim \lambda u $,
for any complex number $\lambda$ which is not zero,
$\lambda \in {\bf C} - \{ 0\} $.
This reduces the space to one complex dimension.
Utilizing this identification, we can take
$\beta /\alpha =z$ as the local complex coordinate
of ${\bf CP}^1$ except in the neighborhood of
$\alpha =0$; a convenient normalization is to take
${\bar \alpha} \alpha = (1+ z\bz )^{-1}$.
(Near $\alpha =0$, we can use $\alpha /\beta$
as the local coordinate.)

There is a natural ${\rm SL(2,{\bf C})}$ action on $u$ given by
$u\rightarrow g u$, $g \in {\rm SL(2, {\bf C})}$. The scalar product of
two $u$'s by $(u_1 u_2) = \epsilon_{AB} u^A_1 u^B_2$, where $u^1 =
\alpha , ~u^2 = \beta$.
(The lower index numbers represent the numberings of u's and the
upper ones represent their components.)
This scalar product is invariant under the ${\rm SL(2, {\bf C})}$ action. The current
correlators (\ref{cp2}) may be written for ${\bf CP}^1$ by writing
$z_{12} = - (\alpha_1 \beta_2 - \alpha_2 \beta_1 )/\alpha_1
\alpha_2 = - (u_1 u_2) /\alpha_1 \alpha_2$. Introducing ${\cal J}$
by $\alpha^2 {\cal J} = J$, we find
\beq
\la {\cal J}^{a_1}(1)
\cdots {\cal J}^{a_n}(n)\ra = -{1\over n\pi^n} \left[ {\Tr(t^{a_1}
\cdots t^{a_n}) \over (u_1 u_2) ~(u_2 u_3) \cdots (u_n u_1)} +{\rm
permutations}\right].
\label{cp5}
\eeq
We can take this expression
as something globally valid on ${\bf CP}^1$, (\ref{cp2}) being the
local version valid in a neighborhood which does not include
$\alpha =0$.

We also note that the variation of the WZW action can be written as
\beqar
\delta {\cal S}_{WZW}&=& - {1\over \pi} \int
d^2 z ~
 \Tr \left(
M^{\dagger -1} \del_\bz M^\dagger \del (M^{\dagger -1} \delta
M^\dagger ) \right)
\nonumber\\
&=& {1\over \pi} \int \Tr ( D_\bz {\cal A}_z M^{\dagger -1} \delta M^\dagger )
= -{1\over \pi} \int \Tr ( {\cal A}_z D_\bz (M^{\dagger -1} \delta M^\dagger ) )\nonumber\\
&=& -{1\over \pi}\int \Tr ( {\cal A}_z \delta A_\bz )
\label{an2.48}
\eeqar
where $D_\bz$ is the covariant derivative in
the adjoint representation,  $D_\bz {{\cal A}_z} = \del_\bz {{\cal
A}_z} + [A_\bz , {{\cal A}_z}]$. ${{\cal A}_z}$ is defined by
\beq
{{\cal A}_z} = M^{\dagger -1} \del_z M^\dagger .
\label{an2.49}
\eeq
Notice that this obeys the equation
\beq
\del_z A_\bz - \del_\bz
{{\cal A}_z} ~+~ [{ {\cal A}_z}, A_\bz] =0 .
\label{an2.50}
\eeq

Putting these considerations aside for the moment and turning to the Yang-Mills
theory, the maximally helicity violating (MHV) tree amplitudes
correspond to the scattering of $n-2$ gluons of negative
helicity and $2$ gluons of positive helicity
(or the other way)
and are given by
\cite{pt}
\beq
{\cal A} (++--\cdots -) =i g^{n-2} ~ (u_I u_J)^4  ~{\Tr
(t^{a_1} t^{a_2}\cdots t^{a_n})\over (u_1u_2) (u_2u_3) \cdots (u_n
u_1)} ,
\label{mg3}
\eeq
where $g$ is the coupling constant. The gluons
are all massless described by null momenta $p_\mu$ with $p^2 =0$.
$u$'s are the spinor momenta of particles given by $p_{A{\dotA}}=
p_\mu (\sigma^\mu )_{A{\dotA}} = u_A \bu_{\dotA}$, where
$\sigma^\mu =(1, {\bf \sigma} )$ and ${\bf \sigma}$ are Pauli
matrices.
The labels $I$ and $J$ refer to the
positive helicity gluons. For simplicity of presentation, all
gluons are taken as
incoming. The expression ${\cal A}$ in (\ref{mg3}) is actually a
subamplitude, the full amplitude is obtained by summing over such
subamplitudes with all noncyclic permutations. This subamplitude
has cyclic symmetry, so we can also sum over all permutations and
divide by $n$. There is also a momentum conservation
$\delta$-function which we have not displayed.

The MHV amplitude (\ref{mg3}), (with momentum conservation inserted),
can now be written as
 \beqar
 {\cal A}(u,\bar{u}) &=& \int \prod_n
d^2 v_n
e^{{i\over 2} v_n \cdot \bu_n }~ {\tilde {\cal A}}(u,v)~~,\nonumber\\
{\tilde {\cal A}}(u,v) &=&  \int d^4x \prod_n \delta (
v_{{\dotA}n} - x_{A{\dotA}}u^A_n) ~i g^{n-2} ~ (u_I u_J)^4  ~{\Tr
(t^{a_1} t^{a_2}\cdots t^{a_n})\over
(u_1u_2) (u_2u_3) \cdots (u_n u_1)}~~.\nonumber\\
 \label{twist2}
 \eeqar
The Fourier-transformed amplitude ${\tilde{\cal A}}$ is
holomorphic in the twistor variable
$Z_\alpha = ( v_{\dotA},u_A)$.
The $\delta$-functions in
${\tilde{\cal A}}$ show that it has support at various points
$u^A_n$ (and corresponding $v$'s) on a line $v_{\dotA} =
x_{A{\dotA}} u^A$. This is a complex line in the space of $Z$'s,
$x_{A{\dotA}}$ specifying the choice of this line. Equation
(\ref{twist2}) was the form used in \cite{witten} to relate these
amplitudes to the topological $B$-model.

The generator of Lorentz transformations
for the $u$'s is given by
\beq
J_{AB} = {1\over 2}\left( u_A {\del \over \del u^B}
+ u_B {\del\over \del u^A}\right) ,
\label{mg4}
\eeq
where $u_A = \epsilon_{AB} u^B$.
The spin operator is given by
$S_\mu =-{\half} \epsilon_{\mu\nu\alpha\beta} J^{\nu\alpha}p^\beta$,
where $J^{\mu\nu}$ is the full Lorentz generator. This works out
to $S_{A \dotA} = J_{AB} u^{B}\bu_\dotA =
s~ p_{A\dotA}$ identifying the helicity as
\beq
s = {1\over 2} u^A {\del \over
\del u^A} .
\label{mg5}
\eeq
Thus $s$ is half the degree of
homogeneity in the $u$'s.

One of the basic observations made in \cite{nair} was that the
subamplitude could be written as
\beqar
{\cal A} (1,2, \cdots, n)
 &=& i g^{n-2} \int d^4x~ d^2\theta_1
d^2\theta_2 d^2\theta_3 d^2\theta_4 \prod_i e^{ip_i\cdot x}\nonumber\\
&&\hskip .1in \times \la A^{a_1}(p_1) A^{a_2}(p_2)\cdots A^{a_n}(p_n)
\ra .
\label{mg9}
\eeqar
In this formula
\beq
A^a (p ) = {\cal J}^a ~\phi (u , \bu ) ,
\label{mg8}
\eeq
where ${\cal J}^a$ is the current of a WZW theory on ${\bf CP}^1$, which 
satisfies the current correlators (\ref{cp5}) and hence
has degree of homogeneity in $u$ equals -2,
and
$\phi (u, \bu )$
is the ${\cal N}=4$ superfield
\beq 
\phi (u,\bu )
= a_{-} + \xi^i a_i + \half \xi^i \xi^j a_{ij} +{1\over 3!} \xi^i
\xi^j \xi^k \epsilon_{ijkl} {\bar a}^l +\xi^1 \xi^2 \xi^3 \xi^4
a_{+} ,
\label{mg7}
\eeq
where $\xi^i = (u\theta )^i = \epsilon_{AB}
u^A \theta^{Bi}$ ($i,j=1,2,\cdots {\cal N}$ are the supersymmetry
indices). We can interpret $a_+$ as the classical value of the
annihilation operator for a positive helicity gluon, $a_-$ as the
annihilation operator for a negative helicity gluon. The
components $a_i , {\bar a}^i$ correspond to four spin-$\half$
particles and $a_{ij}$ correspond to six spin-zero particles, in
agreement with the particle content of ${\cal N}=4$ theory
\footnote{ The particle content of ${\cal N}=3$ is the same as
that of ${\cal N}=4$, but the MHV amplitude
is more naturally expressed for ${\cal N}=4$. See the paper of Rosly and
Selivanov \cite{others} for a comparison.}.
Notice
that the assignment of helicity is consistent with equation
(\ref{mg5}).
The expectation value in (\ref{mg9}) is taken as in the WZW
theory, which means that we can use (\ref{cp5}).
Formula
(\ref{mg9}) also includes the momentum conservation
$\delta$-function; it is generated by the $x$-integration.
Further, it includes similar amplitudes for the superpartners,
namely, the fermions and the scalars, though these do 
not contribute to the $classical$ scattering of
gluons.

We now want to carry out one more step of consolidation by defining
an action for these amplitudes. The WZW action is defined in two dimensions.
There are three independent components for a null momentum vector.
So a slight generalization is needed. The Lorentz-invariant
volume element in momentum space can be written as
\beqar
d\mu (p) &=& {d^3 p \over 2p_0}\nonumber\\
&=& {1\over 4i} [(udu) d^2\bu -  (\bu d\bu ) d^2u ]\nonumber\\
&=&  {1\over 2 }({\bar \alpha}\alpha ) d ({\bar\alpha}\alpha )
{dz d\bz \over (-2i)}.
\label{mg10}
\eeqar
In terms of the spinor $u$, we have still kept the identification
of the
local ${\bf CP}^1$ coordinate,
in the coordinate patch we are working with, as
$z = \beta /\alpha$.

Let ${\cal S}_{WZW}$ be the action for the WZW theory defined in
(\ref{cp1}).
But, one can easily embed it on a more general space by
generalizing $A_{\bar{z}}(z,\bar{z})$ to
$A_{\bar{z}}(z,\bar{z},\cdots)$ in some proper way.
We now use a specific form for $A_\bz$ given
by
\beq 
A_\bz (z,\bar{z}; x^{\mu}, \theta^{{A}i})  
= \pi \int d({\bar \alpha}\alpha ) { {\bar \alpha}\over 
2 \alpha} ~{\widetilde A}\label{mg10a} 
\eeq
The WZW action is expressed in terms of this potential as
($\widetilde{A}(n)$ denotes ${\widetilde A} (u_n,\bar{u}_n,
x^{\mu},\theta^{{A}i})$)
\beq
{\cal S} [{\widetilde A}]
= -\sum_{n=2}^\infty {1\over n} \int d\mu (p_1) \cdots d\mu (p_n)
\Tr \left[{ {\widetilde A}(1) \cdots {\widetilde A}(n) \over
(u_1u_2) (u_2u_3)\cdots (u_n u_1)} \right] .
\label{mg11}
\eeq
We choose ${\widetilde A}$ to be given by
\beqar
{\widetilde A}
(u_n,\bar{u}_n, x^{\mu},\theta^{{A}i})
&=& t^a \left(a^a_{-} + \xi^i a^a_i + \half \xi^i \xi^j a^a_{ij}
+{1\over 3!} \xi^i \xi^j \xi^k \epsilon_{ijkl} {\bar a}^{al}
+\xi^1 \xi^2 \xi^3 \xi^4 a^a_{+} \right)\nonumber\\
&&\hskip .4in \times~e^{ip\cdot x}.
\label{mg12}
\eeqar
Notice that
a part of this field is the same as the superfield $\phi$ of
(\ref{mg7}) except for the extra color index we added. The
scattering amplitude can now be written as
\beqar
{\cal A}
&=&\left[ \left({\delta \over \delta a^{a_1}(p_1)} \right) \cdots
\left( {\delta \over \delta a^{a_n}(p_n)} \right) \exp\left( i
\Gamma [{\tilde A}] \right)
\right]_{{\widetilde A}=0},
\nonumber\\
\Gamma [{\tilde A}] &=& \int d^4x~ d^2\theta_1
d^2\theta_2 d^2\theta_3 d^2\theta_4  {1\over g^2}
{\cal S}[ g{\tilde A}] .
\label{mg13}
\eeqar
If we consider $n$ external gluons one must consider two positive
helicity and $n-2$ negative gluons in order to saturate the
Grassmann integration; in this construction, ${\cal N}=4$ is
crucial to get the right MHV amplitude (\ref{mg3}), and moreover
the vanishing of amplitudes $ {\cal A} (--\cdots -)=0,~{\cal A}
(+--\cdots -)=0$ is automatically satisfied.

Let us now recall that the
$S$-matrix can be expressed in terms of the action as
follows. Let $\Gamma (\vf )$ denote the effective quantum
action of a set of fields, generically denoted by $\vf$.
The quantum equations of motion are the critical points of $\Gamma$ defined
by
\beq
{\delta \Gamma \over \delta \vf } =0 .
\label{smatrix1}
\eeq
The functional which gives the $S$-matrix is then given by
\beq
{\cal F} = \exp \left( i \Gamma \right) \Bigr]_{{\delta\Gamma \over \delta \vf}=0} .
\label{smatrix2}
\eeq
The solutions of the equations of motion depend on a number of free parameters,
which define the phase space of the theory;
the $S$-matrix is a functional of this free data in the solutions.
Thus, for example, in perturbation theory, the solution is obtained
as an expansion around the free field
$\vf = \sum_k a_k u_k(x) + a^*_k u^*_k(x)$, where $u_k (x)$ are plane
wave modes.
The free data are the mode coefficients $a_k ,~a^*_k$.
The amplitude for a process $k_1, k_2, \cdots \rightarrow p_1 , p_2, \cdots$
is given by
\beq
{\cal A} = \Biggl[ {\delta \over \delta a_{k_1}} {\delta \over \delta a_{k_1}}
\cdots {\delta \over \delta a^*_{p_1}} {\delta \over \delta a^*_{p_2}} \cdots
{\cal F} \Biggr]_{a_k = a^*_k =0}.
\label{smatrix3}
\eeq
In the classical theory, we can use the
classical action
${\cal S}_{cl}$
in place
of $\Gamma$.

Notice that the expression (\ref{mg13}) is very similar to
(\ref{smatrix3}). In fact, if we identify
$\int d^4 x d^8\theta
~{\cal S} [{\widetilde A}]$ in (\ref{mg13})
as some sort of classical action for the theory, this is exactly
the expected expression. We shall see below how this can emerge
from the constraints of ${\cal N} =4$ Yang-Mills theory.

\section{A solution to the constraints of ${\cal N} =4$
Yang-Mills theory}

In the ${\cal N}=4$ super Yang-Mills theory,
superspace is described by
$(x^\mu , \theta^{Ai}, {\bar\theta}^{\dotA}_i)$
and we introduce
the standard spinorial derivatives
\beq
D_{A i} = {\del \over \del \theta^{A i}}
+i (\sigma^\mu )_{A {\dotA}} {\bar \theta}^{{\dotA}}_i
{\del \over \del x^\mu},
\hskip .2in
{D}^i_{{\dotA}} =- {\del \over \del {\bar\theta}^{{\dotA}}_i}
-i \theta^{A i}(\sigma^\mu )_{A{\dotA}}
{\del \over \del x^\mu} .
\label{cs19} 
\eeq
We also have
the usual derivative $\del /\del x^\mu$.
We then introduce gauge potentials $A_{A i}, ~ \bA^i_{{\dotA}},~
A_\mu$, which are functions of $x^\mu, \theta^{A i},
{\bar \theta}^{{\dotA}}_i$, corresponding to these
derivatives. Generally speaking this will give too many degrees of freedom
and one has to impose constraints which reduce them to the required number
of fields for the chosen value of ${\cal N}$.
For ${\cal N}=4$, the constraints are
\beqar
F_{A i B j}+ F_{B i  A j} &=& 0 ,
\nonumber\\
F^{ij}_{{\dotA}{\dotB}}+ F^{ij}_{{\dotB}{\dotA}}
&=& 0 ,
\label{cs20}\\
F^j_{A i {\dotB}} &=& 0
\nonumber
\eeqar
along with a subsidiary
condition
\beq
W_{ij} = \half \epsilon_{ijkl} {\overline W}^{kl} ,
\label{cs21}
\eeq
where $F_{A iB j} = \epsilon_{AB} W_{ij}$,
$F^{ij}_{\dotA \dotB} = \epsilon_{\dotA \dotB} {\overline W}^{ij}$.
 These
constraints have long been known to be rather stringent and lead
to the equations of motion via the Bianchi identity
\cite{sohnius}. This property shows that the second order
classical equations of motion of the theory are equivalent to a
set of first order equations in an appropriate superspace,
suggesting a certain integrability for the ${\cal N} =4$
Yang-Mills theory \cite{witten2}.

We have seen that the MHV amplitudes have a natural interpetation
in twistor space where there are additional bosonic variables.
This leads to a possible strategy for solving  the constraints.
We will first write them in a larger space which is, more
or less, a variant of
supertwistor space. We will then do a gauge transformation
(depending on the additional variables) to eliminate some
of the usual gauge potentials. In the new gauge, the solution to the
constraints is simpler. Such a method has been used to construct
superfields for ${\cal N} =2$ Yang-Mills theory;
that construction was based on harmonic superspace, which is a
close relative of twistor space \cite{harmonic}.

We start by introducing a complex spinor $u^A$.
(This time we are not thinking of it as a spinor momentum, not yet.)
The complex conjugate of $u^A$ transforms as a dotted spinor,
$\bu_{\dotA} = (u^A)^*$.
So, to get something that transforms
in a similar way to $u^A$,
we introduce a vector $K_{A\dotA}$ and write
$\bw^A = K^{A\dotA} \bu_{\dotA}$.
Thus for a fixed choice of $K$, $\bw^A$ has the same information
as the conjugate of $u^A$. Using these variables, we can take
combinations of the derivatives on superspace
for the undotted sector
as
\beq
D^+_i =  u^A D_{Ai}, \hskip .2in
D^-_i = - \bw^{A} D_{Ai} .
\label{con1}
\eeq
(We will take $K$ such that the scalar product
$(\bw u)$
is not zero.)
We also have similar combinations for the gauge potentials.
The constraints of the ${\cal N} =4$ theory can now
be written as
\beqar
&& F^{++}_{ij} = F^{+-}_{ij}+ F^{-+}_{ij} =
F^{--}_{ij} = 0
\nonumber\\
&&F^{~ij}_{\dotA \dotB} + F^{~ij}_{\dotB \dotA} =0
\label{con2}\\
&&F^{\pm ~j}_{i~~\dotB} = 0.
\nonumber
\eeqar
The components which are not zero are
$F^{+-}_{ij} = (u \bw ) W_{ij}$,
$F^{ij}_{\dotA \dotB} = \epsilon_{\dotA \dotB} {\overline
W}^{ij}$. 

Let
${\D}^+_i, \D^-_i$, $\D^i_{\dotA}$
denote the gauged versions of the spinorial derivatives, 
${\D} = D +A$, with 
the gauge potentials $ A^+_i = 
u^A A_{Ai},~A^-_i =
- \bw^{A} A_{Ai} $ and $A_{\dot{A}i}$, respectively.
We also introduce the additional derivatives
\beqar
&&D^{++} = u^A {\del \over \del \bw^A},
\hskip .2in D^{--} = - \bw^A {\del \over \del u^A},
\nonumber\\
&&D^0 =  \left( u^A{\del \over \del u^A} - \bw^A {\del \over \del
\bw^A}\right).
\label{con3}
\eeqar
Notice that $D_0$ is a charge operator, assigning $+1$ charge to
$u^A$ and $-1$ charge to $\bw^A$. The superscripts in
(\ref{con1}), (\ref{con3}) indicate the value of this charge for
each of the derivatives.

The constraints of the theory can now be displayed as
\beq
\{ \D^{+}_{i}, \D^{+}_{j}\} = 0
\label{con4}
\eeq
\beq [\D^{++}, \D^+_i] = 0
\label{con5}
\eeq
\beq [\D^{--}, \D^+_i ] =\D^-_i
\label{con6}
\eeq
\beq [\D^{++},
\D^{--}] =  - ~D_0
\label{con7}
\eeq
\beqar
&&\{ \D^{+}_{i}, \D^{-}_{j}\}  + \{ \D^{-}_{i}, \D^{+}_{j}\} =  0
\nonumber\\
&&\{ \D^{-}_{i}, \D^{-}_{j}\} = 0
\nonumber\\
&&{}[ \D^{++} , \D^{-}_i ] = - \D^{+}_i
\label{con8}\\
&&{}[ \D^{--}, \D^{-}_i ] =0
\nonumber
\eeqar
\beqar
&& [ \D^{++}, \D^i_{\dotA} ] =0
\nonumber\\
&&{}[ \D^{--} , \D^i_{\dotA}] =0
\label{con9}
\eeqar
\beq
\{\D^{+}_{i},
D^{j}_{\dotA}\}=\delta^{j}_{i}  u^A
\D_{A{\dotA}}
\hskip .2in
\{ \D^{-}_{i},
D^{j}_{\dotA}\}= - \delta^{j}_{i} \bw^A
\D_{A{\dotA}}
\label{con10}
\eeq
\beq
\{ {\cal D}^i_{\dotA} , {\cal D}^j_{\dotB} \} + \{ {\cal D}^i_{\dotB} , {\cal D}^j_{\dotA} \} 
=0\label{con10a}
\eeq
Even though we have written the
gauged versions
$\D^{\pm \pm}$,
the gauge potentials
$A^{++}, ~A^{--}$
are zero at this stage; these constraints are
thus equivalent to the previous constraints (\ref{con2}). Further,
even though we introduced $u^A$, $\bw^A$, the constraints do not
depend on all components of these spinors. The constraints are
homogeneous (of different degrees) and so, one of the components,
say $\alpha$ (and ${\bar \alpha}$) can be factored out.

So far, the introduction of the additional variables and the potentials
is really a meaningless redundancy since their potentials are
zero. However, we now notice that, because of the constraints
(\ref{con4}),  $A^{+}_i$ is of the form
$ -D^+_i g g^{-1}$,
for some matrix $g$ (which is generally not unitary).
The matrix $g$ is in general a function of $x^\mu , \theta^{Ai},
{\bar\theta}^{\dotA}_i$ and the new coordinates $u^A, \bw^A$.
(If it did not depend on $u^A, \bw^A$, $W_{ij}$ would be zero.)
This property of
$A^+_i$ suggests that we can make a
``gauge transformation'' using $g$ and eliminate it. When this is
done, the potentials
$A^{\pm \pm}$
are no longer zero, rather
$A^{++} = g^{-1} D^{++} g$.
In this
new gauge $A^+_i =0$, the constraints (\ref{con4}) to (\ref{con8})
become
\beq
D^{+}_i A^{++} =0
\label{con11}
\eeq
\beq 
A^{-}_i = - D^{+}_i A^{--} \label{con12} 
\eeq
\beq
D^{++} A^{--} - D^{--} A^{++} + [ A^{++}, A^{--} ] =0
\label{con13}
\eeq
\beq
\begin{array}{c}
D^{+}_i A^{-}_j + D^{+}_j A^{-}_i =0
\\
D^{-}_i A^{-}_j + D^{-}_j A^{-}_i +\{ A^{-}_i , A^{-}_j \}=0
\\
D^{++} A^{-}_i - D^{-}_i A^{++} +[ A^{++}, A^{-}_i]=0
\\
D^{--} A^{-}_i - D^{-}_i A^{--} +[ A^{--}, A^{-}_i]=0
\\
\end{array}
\label{con14}
\eeq
In addition to these,
we still have the constraints (\ref{con9}) and (\ref{con10})
as well as (\ref{con10a}) or $F^{ij}_{\dotA \dotB} +
({\dotA} \leftrightarrow {\dotB})
 =0$
in (\ref{con2}).

These equations show how we can obtain a
solution to the theory.
We can start with $A^{++}$ as the
given quantity.
It must be chosen such that it satisfies an analyticity
condition (\ref{con11})
\footnote{This term came from the definition of analyticity in
harmonic superspace \cite{harmonic}.}. Equation (\ref{con13}) then defines
$A^{--}$. Given $A^{--}$ we can use (\ref{con12}) to obtain
$A^{-}_i = - D^{+}_i A^{--}$.
This will give us both $A^{+}_i$ (which is zero)
and $A^{-}_i$; one can even transform back to the original gauge,
if it is convenient.
To show that this is indeed a solution, we must also check the
constraints (\ref{con14}) using
$A^{-}_i = - D^{+}_i A^{--}$.
This
can be done in a straightforward way.

We have thus solved half of the constraints; we must now
consider the dotted sector and the mixed constraints
(\ref{con9}).
Having obtained $A^{\pm}_i$, we can, in principle,
transform them back to the original gauge and take
conjugates to get $A^i_{\dotA}$.
This will take care of the constraints
$F^{ij}_{\dotA \dotB} +
({\dotA} \leftrightarrow {\dotB})
=0$. The constraints (\ref{con10}) can be taken as the definition
of $A_{A \dotA}$. The only difficulty is with the constraints
(\ref{con9}). This constraint reads
\beq
D^{\pm \pm} A^i_{\dotA} -
D^i_{\dotA} A^{\pm \pm} +[ A^{\pm\pm} , A^i_{\dotA}] =0 .
\label{con15}
\eeq
We do not have a way to deal with this in
generality, but we notice that a particular solution may be
obtained by setting
${D}^i_{\dotA} A^{++}$ or ${D}^i_{\dotA} A^{--}$
to zero. This imposes
a chirality condition on $A^{++}$ (and via (\ref{con13}) on
$A^{--}$). Thus for our special solution
we have
\beq
{D}^i_{\dotA} A^{++}=0~.
\label{con16}
\eeq
What we have shown is that if we find an $A^{++}$ obeying the
analyticity condition (\ref{con11}) and the chirality condition
(\ref{con16}), then we can find a solution to the constraints of
the ${\cal N} =4$ theory. The only nontrivial condition is
equation (\ref{con13}).

We now turn to the solution of (\ref{con11}), (\ref{con16})
and (\ref{con13}).
The solution to the
chirality condition (\ref{con16}) is well known: $A^{++}$ must depend
on ${\bar\theta}^{\dotA}_i$ only through the combination
$y^\mu = x^\mu -i {\bar\theta}^{\dotA}_i \theta^{Ai}
(\sigma^\mu )_{A{\dotA}}$
\footnote{One might consider $y^\mu = x^\mu -i
{\bar\theta}^{\dotA}_i \theta^{Ai} (\sigma^\mu )_{A{\dotA}} +F$
with an arbitrary function $F=F(\theta^{A i}, u^A,
\bar{\omega}^A)$ such as $\theta^{Ai} (\sigma^{\mu})_{A \dot{A}}
\bar{u}_{\dot{A}}$, but there is no combination satisfying the condition of being a
singlet for the supersymmetry indices $i$ and having vanishing $D^0$ charge
.}.
We will look for solutions of the form
$A^{++}=A_p \exp (i p\cdot y )$;
the
analyticity condition (\ref{con11}) then tells us that
\beq
\left[ u^A {\del A_p \over \del \theta^{Ai}}
- 2
u^A (\sigma \cdot p )_{A{\dotA}} {\bar \theta}^{\dotA}_i A_p
\right] =0 .
\label{con17}
\eeq
Since the first term does not have a
factor of ${\bar\theta}^{\dotA}_i$, we get a nonzero solution only
if
\beqar
u^A {\del A_p \over \del \theta^{Ai}} &=& 0
\nonumber\\
u^A (\sigma \cdot p )_{A{\dotA}} &=& 0 .
\label{con18}
\eeqar
The first equation tells us that $A_p$ must depend on
$\theta^{Ai}$ only through $\xi^i = u_A \theta^{Ai}$, so that we can
write
\beq
A_p =   t^a
\left(a^a_{-} + \xi^i a^a_i + \half \xi^i \xi^j a^a_{ij}
+{1\over 3!}
\xi^i \xi^j \xi^k \epsilon_{ijkl} {\bar a}^{al}
+\xi^1 \xi^2 \xi^3 \xi^4 a^a_{+} \right) ,
\label{con19}
\eeq
where the coefficients $a^a_{\pm}, ~a^a_i, ~a^a_{ij}, ~{\bar a}^{al}$
are arbitrary functions of $p$
\footnote{ There may exist $(\bw u)$ dependence, but this
would not have an important role in the dynamics with a fixed
choice of $K^{A \dot{A}}$.}.
Notice that we have essentially recovered the superfield of
(\ref{mg12}) from the
(special)
solution of the constraints of the ${\cal N}=4$ theory,
except for the appearance of $y^{\mu}$, instead of $x^{\mu}$, in
the plane wave part $\exp (i p\cdot y )$. (It is immaterial that
$y^\mu$ appears rather than $x^\mu$ since we will be integrating over
$x^\mu$ anyway.)

The second condition in (\ref{con18}) shows that $p_\mu$ must be
a null vector. Thus the solution must be on-shell, as we knew it would be
from the general statement that the constraints put the ${\cal N}=4$ theory
on shell.
$u^a$ is an eigenvector of $\sigma\cdot p$ with zero
eigenvalue. Since $p_\mu$ is real, we can write
$p_{A{\dotA}} = u_A \bu_{\dotA}$.

For a general solution to (\ref{con11}) and (\ref{con16}) we can
do a superposition by integrating over the null momenta. But
recall that $u^A$ was part of the space, so we do not have the
full freedom of integration. All of our constraints really depend
only on the ${\bf CP}^1$ subspace whose local coordinate we have
taken as
$z = \beta/\alpha$;
there is freedom to divide out by an appropriate number of
$\alpha , ~{\bar\alpha}$ because there is a balance of charges. So
what can be freely integrated over is just the $\alpha$ part. We
choose this measure to be consistent with Lorentz invariance; this
brings us to the combination $A_\bz$ given in (\ref{mg10a}) in
terms of an integral over
${\widetilde A}=A_p \exp (i p\cdot y )$.

The final equation to be solved, namely (\ref{con13}),  is
now straightforward.
First of all, we express it in local coordinates.
While $u^A$ and $(u^A)^*$ define the usual complex coordinates in terms of
which we can get the local coordinates $z, \bz$,
we had to introduce a vector $K$ to obtain Lorentz invariant
contractions and
to
define the derivatives
$D^{\pm \pm}$.
If we choose $K^A_{~\dotA} =
\delta^A_{~\dotA}$, this will correspond to the usual description
of ${\bf CP}^1$ where we use $u^A$ and $\bu^A$. (This corresponds
to the use of a particular frame to define the derivatives $D^{\pm
\pm}$, but our final results will be Lorentz invariant.) We now
define a set of local gauge potentials $A_\bz , {\cal A}_z$ by
\beqar
A^{++} &=& {\alpha \over {\bar\alpha}} (1+z \bz ) ~A_\bz ,
\nonumber\\
A^{--}&=& {{\bar\alpha}\over \alpha} (1+z\bz ) ~
{\cal A}_z ,
\label{con20}
\eeqar
The substitution of these into equation (\ref{con13}) transforms it into
\beq
\del_\bz
{\cal A}_z
 - \del_z A_\bz + [ A_\bz ,
{\cal A}_z
 ] =0.
 \label{con21}
\eeq
Basically, this takes us to the equation (\ref{an2.50}).
What we have shown is that solving 
(\ref{con13}) is equivalent to solving (\ref{con21}).
Given a solution of (\ref{con21}), we can obtain
a solution to (\ref{con13}) by using (\ref{con20}).

Equation (\ref{con21}) can be solved for ${\cal A}_z$ in terms of $A_\bz$;
the latter is arbitrary except for the analyticity condition (\ref{con11}).
Notice that if we substitute (\ref{con20}) into
(\ref{con11}), the prefactor $({\alpha / {\bar\alpha}}) (1+z \bz )$ drops out; we
can also factor out $\alpha$ from $u^A$ (which is equivalent to
writing $u^1 =1, ~u^2 =z$. We thus obtain the
same conditions
(\ref{con18}) for $A_\bz$ with $(1, z)$ in place of $u^A$.
(This is in accordance with our earlier comment on dividing out 
$\alpha, ~{\bar \alpha}$.)
It can be solved for $A_\bz$ as before, giving a function of the 
$z$'s and the momentum $p$; the value of
$\alpha, ~ {\bar\alpha}$ given by the momentum can then be used
to go back to the full $u^A$.
We see that $A_\bz$ is given by a superposition of fields of the form
$A_p e^{ip\cdot y}$ with $A_p$ given by (\ref{con19}).
We now take it to be given by
\beqar
A_\bz &=&  \pi \int d({\bar \alpha}\alpha )
{{\bar \alpha}\over 2 \alpha} A_p
e^{ip\cdot y}
\nonumber\\
&=&
\pi \int d({\bar \alpha}\alpha )
{{\bar \alpha}\over 2 \alpha}
{\widetilde A}.
\label{con20a}
\eeqar
This is essentially (\ref{mg10a}), but we have obtained it
as a solution of the constraints. 
(It should be emphasized that since the coefficients
$a^a_-, ~a^a_i$, etc., can also be functions of $p$, there is no
loss of generality in taking the particular form (\ref{con20a}). In other words, it is
simply the choice for which the coefficients can be interpreted as the 
properly normalized annihilation
amplitudes.)

The key issue is thus the solution of
(\ref{con21}) for ${\cal A}_z$. But rather than discussing ${\cal A}_z$ in
its own right, we shall now turn to the $S$-matrix. Since we are
looking for tree-level amplitudes at this point, what we need, in
the spirit of (\ref{smatrix2}), is a classical action. The basic
equation of motion for us is (\ref{con13}) or (\ref{con21}). We need an action,
which for any given
$A_{\bar{z}}$,
gives the equation (\ref{con21}) for
${\cal A}_z$.
This action, not surprisingly, is a variant of the WZW action
in the holomorphically extended superspace $(x^{\mu},
\theta^{{A}i}, z,\bar{z})$
and is
\beq {\cal S}  = - k \int d {\bf X} ~{\cal S}_{WZW}(U) +{k\over \pi} \int d
{\bf X} d^2 z ~\Tr (
 A_\bz \del_z U U^{-1}).
 \label{con22}
\eeq
Here $k$ is a normalization
constant which can be thought of as
the level number of the WZW action and $d {\bf X}=d^4 x d^2\theta_1
d^2\theta_2 d^2\theta_3 d^2\theta_4$.
The equation of motion can be obtained by varying the matrix field
$U$ and is identical to (\ref{con21}) with
${\cal A}_z = - \del_z U U^{-1}$.
 Further, if we write $A_\bz = M^{\dagger -1} \del_\bz
M^\dagger$, for some matrix $M^\dagger$, the solution to
(\ref{con21}) is evidently $U= M^{\dagger -1}$. We can now use the
Polyakov-Wiegman
identity to write
\beqar
-{\cal S}_{WZW}(U)
+{1\over \pi} \int  d^2 z ~\Tr ( A_\bz \del_z U U^{-1})
&=& -{\cal S}_{WZW} (M^\dagger U)~+~ {\cal S}_{WZW} (M^\dagger )\nonumber\\
&=& {\cal S}_{WZW} (M^\dagger )\nonumber\\
&=& {\cal S}_{WZW} (A_\bz ).
\label{con23}
\eeqar
This tells us that the action (\ref{con22}), which leads to the
required equation of motion (\ref{con21}), when evaluated on solutions
of that equation is given by the WZW action of (\ref{cp1})
(with the additional integration with the measure $d {\bf X}$).
All we have to do at this point is to substitute the form of
$A_\bz$ given by (\ref{con20a}) to obtain the $S$-matrix
amplitudes,
following the general formula (\ref{smatrix3}).
Evidently, we have recovered the formula (\ref{mg13})
\footnote{ There are additional $\bar{\theta}^{\dot{A}}_i
{\theta}^{{A} i}$-terms from $\exp (i p \cdot y)$, since
$y^\mu = x^\mu -i
{\bar\theta}^{\dotA}_i \theta^{Ai} (\sigma^\mu )_{A{\dotA}} $,
but the tree amplitudes are the
same as those of (\ref{mg12}) since ${\cal A}(--\cdots-)\sim
(\bar{\theta})^8\vert_{\bar{\theta}^{\dot{A}}_i=0}=0,~{\cal
A}(+-\cdots-)\sim
(\bar{\theta})^4\vert_{\bar{\theta}^{\dot{A}}_i=0}=0$.}.
Notice also that our final formula
(\ref{mg13})
involves only
Lorentz-invariant scalar products; thus, the choice of the vector
$K$ is irrelevant. (We can recover the coupling constant by the
standard scaling
${\widetilde A} \rightarrow g {\widetilde A}$.
But the overall
normalization of the action
$k$
is not given by the equations of
motion. This is always the case classically. Thus there is one
constant in the amplitudes which is not determined by our
argument. This is basically Planck's constant.)

We also note that
if we introduce two more variables $\zeta_{\dotA}$  and write
combinations like ${\bar D}^{i+} = \zeta^{\dotA} D^i_{\dotA}$,
then we can obtain similar results for the opposite
``handedness'', with almost all positive helicity gluons ${\cal
A}(++\cdots +)=0,~ {\cal A}(-++\cdots +)=0$, and ${\cal
A}(--++\cdots +)$ by exchanging the undotted-sector with
dotted-sector. Furthermore, if we were to introduce both $u_A$ and
$\zeta_{\dotA}$, then
we are naturally led to a
${\bf CP}^1 \times {\bf CP}^1$
structure. This has occurred before in connection with the ${\cal
N}=4$ theory, for example, the paper of Rosly and Schwartz in
\cite{harmonic} as well as \cite{witten, string}. It would be
interesting to utilize this structure as well as the Chern-Simons
theory description to eliminate the condition (\ref{con16}).

\section{The non-MHV amplitudes}

So far our analysis is restricted to the MHV amplitudes. In fact,
we see that, once we make the simplifying assumption of
$D^i_{\dotA} A^{++}=0$, we are restricted to the MHV amplitudes.
The proper way to proceed would thus be to relax this condition
and see how the solution to the constraints would change. However,
this is rather difficult; our derivation is limited to the MHV
amplitudes. As mentioned in the introduction, a suggestion was
made in \cite{witten3} that one could simplify the calculation of
the non-MHV amplitudes by using MHV vertices and then connecting
them via propagators, analogous to Wick contractions in standard
perturbative field theory. While we do not have an independent
justification or derivation of this result, we note that there is
an elegant way to incorporate it in our formalism.

The Wick contraction operator for two gluons is given by
\beq
{\hat W} = \exp \left[ - \int_{x,y} D(x,y) {\delta \over \delta a^a_- (x) }{\delta \over \delta a^a_+ (y)}
\right]
\label{nonMHV1}
\eeq
with the propagator $D(x,y)$ which is the inverse of $p^2$.
Consider the functional for the $S$-matrix defined by
\beq
{\cal
F} = {\hat W} ~ \exp (i \Gamma [{\widetilde A}])
\label{non-MHV2}
\eeq
where
$\Gamma [{\widetilde A}]$
is given in (\ref{mg13}). Consider the
application of this to two vertices, resulting again in a tree
diagram. First of all, to include propagators, we need the
off-shell continuation of the amplitudes, at least for the gluon
which is replaced by the propagator. This will be assumed to be
done as in \cite{witten3}. The prescription is the following. If
$p_\mu$ is the off-shell momentum, the corresponding spinor
momentum in the MHV vertex will be taken as $u_A =
p_{A{\dotA}}\xi^{\dotA}$, where $\xi^{\dotA}$ is a fixed spinor,
taken to be the same for all off-shell lines in a diagram.
Secondly, the individual MHV amplitudes have a color structure of
the form $\Tr (t^{a_1} \cdots t^{a_n})$. Since $U(1)$'s decouple
from the theory, we may extend the range of the indices $a_1,
~a_2$, etc., to include a $U(1)$ direction as well. We will take
the $t^a$'s to be normalized so that we have the completeness
relation $(t^a)_{ij} (t^a)_{kl} = \delta_{jk} \delta_{il}$. Then
the contractions preserve the color structure $\Tr (t^{a_1} \cdots
t^{a_n})$ with cyclic ordering of the external lines from the
individual vertices.

Using (\ref{non-MHV2}), we may calculate the subamplitude
${\cal A} (1_- 2_- 3_- 4_+ 5_+ 6_+ )$ for the scattering of six gluons. We find
\beqar
{\cal A} (1_- 2_- 3_- 4_+ 5_+ 6_+ ) &=& {\cal A}(4_+ 5_+ 6_+ 1_- k_- )
D^{(1)}_{kl} {\cal A} (l_+ 2_- 3_-) \nonumber\\
&& + {\cal A}( 3_- 4_+ 5_+ 6_+  k_-)
D^{(2)}_{kl} {\cal A}(l_+ 1_- 2_-)\nonumber\\
&&+ {\cal A}(6_+ 1_- k_-) D^{(3)}_{kl} {\cal A} (l_+ 2_- 3_- 4_+ 5_+)\label{non-MHV3}\\
&&+ {\cal A}(3_- 4_+  k_-) D^{(4)}_{kl} {\cal A} (l_+ 5_+ 6_+ 1_- 2_-)\nonumber\\
&&+ {\cal A}(3_- 4_+ 5_+  k_-) D^{(5)}_{kl} {\cal A} (l_+  6_+ 1_- 2_-)\nonumber\\
&&+{\cal A}(5_+ 6_+ 1_- k_-) D^{(6)}_{kl} {\cal A} (l_+ 2_- 3_- 4_+),
\nonumber
\eeqar
where the $D$'s are given by
\beqar
D^{(1)}_{kl} = (p_2 +p_3 )^{-2}, &\hskip .5in& D^{(2)}_{kl} = (p_1 +p_2 )^{-2},
\nonumber\\
D^{(3)}_{kl} = (p_6 +p_1 )^{-2}, &\hskip .5in& D^{(4)}_{kl} = (p_3 +p_4 )^{-2},
\label{non-MHV4}\\
D^{(5)}_{kl} = (p_3 +p_4+p_5 )^{-2}, &\hskip .5in& D^{(6)}_{kl} = (p_2 +p_3 +p_4)^{-2}.
\nonumber
\eeqar
This result agrees with the general prescription given in \cite{witten3}.
The general formula (\ref{non-MHV2}) can also generate loop diagrams. It is not
entirely clear to us at this point whether they are identical to the
one-loop amplitudes of the ${\cal N}=4$ Yang-Mills theory, although the
resulting amplitudes are very similar to the recent suggestion
in \cite{spence}.
\vskip .2in\noindent
{\bf Acknowledgments}
\vskip .2in\noindent
VPN thanks the Abdus Salam International Centre for Theoretical Physics, Trieste,
for hospitality during the course of this work.
VPN was supported by the National Science Foundation grant number
PHY-0244873 and by a PSC-CUNY award.
MIP was supported by the Korean Research Foundation Grant
KRF-2002-070-C00022.

\end{document}